\documentclass[aps,pre,twocolumn,superscriptaddress]{revtex4-2}
\usepackage{mathrsfs}
\usepackage{physics}
\usepackage{hyperref}
\usepackage{bm}
\usepackage{url} 
\usepackage{graphicx}
\usepackage{subfigure}
\usepackage{xcolor}
\usepackage{bm}
\usepackage{bbm}
\usepackage{stix}
\usepackage{soul}
\usepackage[normalem]{ulem}

\newcommand{\ba}{\begin{equation}\begin{aligned}}
\newcommand{\ea}{\end{aligned}\end{equation}}

\makeatletter

\begin{document}
\title{Stochastic thermodynamic bounds on logical circuit operation}
\author{Phillip Helms}
\thanks{These authors contributed equally to this work.}
\affiliation{Department of Chemistry, University of California, Berkeley, California 94720, USA}
\affiliation{Chemical Sciences Division, Lawrence Berkeley National Laboratory, Berkeley, California 94720, USA}
\author{Songela W. Chen}
\thanks{These authors contributed equally to this work.}
\affiliation{Department of Chemistry, University of California, Berkeley, California 94720, USA}
\author{David T. Limmer}
\email{dlimmer@berkeley.edu}
\affiliation{Department of Chemistry, University of California, Berkeley, California 94720, USA}
\affiliation{Chemical Sciences Division, Lawrence Berkeley National Laboratory, Berkeley, California 94720, USA}
\affiliation{Materials Sciences Division, Lawrence Berkeley National Laboratory, Berkeley, California 94720, USA}
\affiliation{Kavli Energy NanoScience Institute, Berkeley, California 94720, USA}

\date{\today}

\begin{abstract}
Using a thermodynamically consistent, mesoscopic model for modern complementary metal-oxide-semiconductor transistors, we study an array of logical circuits and explore how their function is constrained by recent thermodynamic uncertainty relations when operating near thermal energies.
For a single NOT gate, we find operating direction-dependent dynamics, and a trade-off between dissipated heat and operation time certainty. 
For a memory storage device, we find an exponential relationship between the memory retention time and energy required to sustain that memory state.
For a clock, we find that the certainty in the cycle time is maximized at biasing voltages near thermal energy, as is the trade-off between this certainty and the heat dissipated per cycle. 
We identify a control mechanism that can increase the cycle time certainty without an offsetting increase in heat dissipation by working at a resonance condition for the clock.
These results provide a framework for assessing thermodynamic costs of realistic computing devices, allowing for circuits to be designed and controlled for thermodynamically optimal operation. 
\end{abstract}
\maketitle

\section{Introduction}

While semiconductor-based computational capacity \cite{moore1965cramming,moore1975progress}
and efficiency \cite{dennard1974design,leiserson2020there,shalf2020future}
has exhibited sustained exponential growth over the past century, 
continued adherence of these trends is being disrupted as 
feature sizes approach atomic length scales and energetic 
scales near those of thermal noise \cite{leiserson2020there,shalf2020future,koomey2010implications}. 
At such small scales, computation has to reconcile with unavoidable noise \cite{johnson1928thermal, nyquist1928thermal}. 
This noisy limit has been termed \textit{thermodynamic computing}
 \cite{conte2019thermodynamic,hylton2021vision} and requires the development of new principles to achieve robust and energy-efficient information processing \cite{markovic2020physics,kaiser2021probabilistic,camsari2019p,wolpert2020thermodynamics,dago_adiabatic_2023}. 
 In this paper, we explore fundamental limitations encountered when computing in this regime 
by showing how the function of realistic logical circuits is bounded by recent thermodynamic uncertainty relations \cite{barato2015thermodynamic,gingrich2016dissipation}. 

Building upon equilibrium thermodynamics-based limits on computing operations, such as Landauer's limit on the cost of bit erasure \cite{landauer1961irreversibility}, stochastic thermodynamics \cite{seifert2012stochastic, wolpert_is_2023} provides a framework for exploring the inherent limits of logical circuit operations on small scales, far from equilibrium. Recent results like fluctuation theorems, thermodynamic uncertainty relations, and speed limits \cite{shiraishi2018speed, ito2018stochastic,macieszczak2018unified,falasco2020dissipation,
ito2020stochastic, kuznets2021dissipation, neri2022universal,horowitz2017proof, gingrich2017inferring, wolpert2019stochastic} can be used to strengthen bounds on computation within the thermodynamic computing regime, provided a physically-consistent, stochastic model. Using a recently developed model for current complementary metal-oxide-semiconductor (CMOS) transistors, \cite{gao2021principles} we study the interplay between accuracy, speed, and heat dissipation of an array of computations performed near thermal energies, locating optimal trade-offs between
thermodynamic and operational costs. 

This paper is organized as follows. In Sec. \ref{model}, we describe the system discussed and the assumptions made. In Sec. \ref{inverter}, we describe the behavior of a NOT gate and how the speed of its operation is weakly constrained by thermodynamics. In Sec. \ref{memory}, we characterize a memory storage device and show it efficiently converted energy into memory preservation. In Sec. \ref{clock}, we extend the discussion to a clock, and in Sec. \ref{control} explore techniques for controlling it to enhance its accuracy without requiring excess energy consumption.

\section{Model} \label{model}
Many conventional engineering approaches for characterizing the effects of noise on circuit
operation rely on assumptions only valid near equilibrium or near specific operating 
conditions \cite{van1970noise, heinen1991unified, rizzoli1988state, weiss1998thermodynamical}, 
guaranteeing neither thermodynamic consistency nor accuracy far from equilibrium \cite{gao2019nonlinear}. 
To provide a more faithful description of stochastic circuits, we require models 
that obey local detailed balance and exhibit shot noise \cite{scholten2006compact}, 
while accurately reproducing known circuit characteristics. 
Recently, several stochastic models for CMOS devices have been proposed \cite{gu2020counting,freitas2021stochastic,gao2021principles}, 
enabling the study of noisy circuits and the associated thermodynamic costs when operating these devices near thermal energies \cite{freitas2022reliability,kuang2022modelling, gopal2022large,yoshino_thermodynamics_2023,peng_modeling_2024}.
Here, we employ one such model \cite{gao2021principles} to study systems of inverters, or logical NOT gates, 
built from single electron tunnel junctions operating within the classical limit \cite{datta1997electronic}, and using a capacitive charging model for the readout voltage. This model meets the three criteria emphasized above and in principle can be parameterized directly from microscopic calculations, providing a link between circuit performance and the underlying materials properties.

As shown in Fig.~\ref{fig:not_gate}a, each inverter contains an N-type and a P-type transistor, each modeled by the band energy of an electron in the transistor $\epsilon_\mathrm{N}\left(V_\mathrm{in}\right)$ and $\epsilon_\mathrm{P}\left(V_\mathrm{in}\right)$, respectively, with energy levels controlled by the inverter's input voltage $V_\mathrm{in}$.
This form of $\epsilon_i$ is valid in the limit of high gate capacitance.  We set $\epsilon_\mathrm{P}=qV_\mathrm{in}$ and  $\epsilon_\mathrm{N}=\frac{3}{2}qV_\mathrm{d}-qV_\mathrm{in}$ to reproduce the characteristic voltage transfer curve of an inverter where $q$ is the unit of charge \cite{gao2021principles}.
The transistors are connected to three electron reservoirs: a source connected to the N-type transistor with reference voltage $V_\mathrm{s}=0$, a drain connected to the P-type transistor with voltage $V_\mathrm{d}>0$ resulting in a cross-voltage, and an output gate connected to both transistors. 
While the source and drain are held fixed, the output gate voltage changes as electrons accumulate in the gate according to $dV_\mathrm{g}/dt=-J_\mathrm{g}(t)/C_\mathrm{g}$, where $J_\mathrm{g}$ is the current of electrons into the gate and $C_\mathrm{g}$ is the gate capacitance. We use $C_\mathrm{g}=10 q/V_\mathrm{T}$ throughout, though 
we have verified nearly identical inverter performance at $C_\mathrm{g}=5 q/V_\mathrm{T}$ and $C_\mathrm{g}=15 q/V_\mathrm{T}$ \cite{datta1997electronic}. 

The system evolves stochastically according to a Markovian master equation 
$\partial_t \mathbf{P}(t)=\mathbf{W} \mathbf{P}(t)$ where $\mathbf{P}$ is 
the configurational probability vector and $\mathbf{W}$ is the stochastic generator, 
with elements $W_{ij}$ specifying the rate at which an electron transitions from state $j$ to $i$
and $P_{i}(t)$ being the probability of being in state $i$ at time $t$.  
The ratio of forward and reverse rates satisfy local detailed balance, $W_{ij}/W_{ji}=e^{-\beta(E_i-E_j)}$, 
where $E_i$ is the energy of a given configuration and 
$\beta=1/k_\mathrm{B} T$ is the inverse temperature defined with Boltzmann's constant $k_\mathrm{B} $
and temperature $T$. The rates are defined using the Fermi-Dirac distribution
where the transition rate of an electron from an electrode $j$ to a transistor $i$ is
$W_{ij}=\Gamma \left(e^{\beta(\epsilon_i-qV_j)}+1\right)^{-1}$
and the reverse is
$W_{ji}=\Gamma -\Gamma\left(e^{\beta(\epsilon_i-qV_j)}+1\right)^{-1}$
where $\Gamma$ 
specifies the timescale for transitions and is physically set by the resistance of the transistor-electrode interface. 
To ensure local detailed balance and avoid the Brillouin paradox, we use the average of the gate voltage $V_j$ before and after each transition \cite{brillouin,freitas2021stochastic}.

We work in units of thermal voltages and times,  $V_\mathrm{T}=k_\mathrm{B}T/q$ and $\beta\hbar$ respectively, and set 
$\Gamma^{-1}=5\beta\hbar$ with $\hbar$ being Planck's constant. We set these constants to ensure the weak coupling limit, in which electron transitions can be treated as discrete hopping events \cite{datta1997electronic}.
For reference, at room temperature $V_\mathrm{T}\approx26 \ \mathrm{mV}$ and 
$\beta\hbar\approx25 \ \mathrm{fs}$. 
In these units, the model inverter is determined  by a specification of the input voltage $V_\mathrm{in}$
and cross-voltage $V_\mathrm{d}$. 
While we first study the behavior of a single inverter, 
by supplying the output voltage of an inverter
as the input voltage to another, 
more complex functionalities can be realized, such as the memory device and 
clock we subsequently consider.

\section{NOT Gate} \label{inverter}

\begin{figure}
\includegraphics[width=8.6 cm]{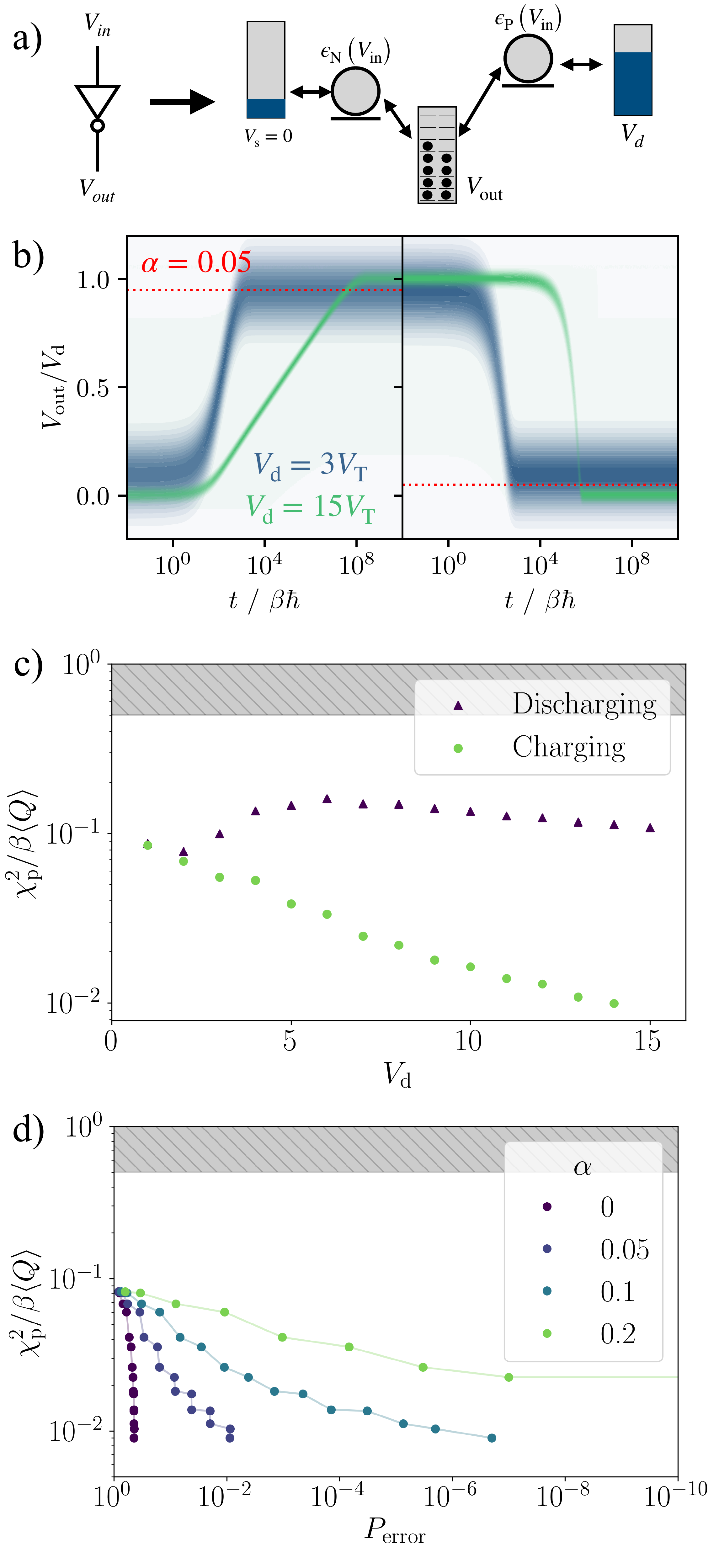}
\caption{
Characterization of the NOT gate.
(a) Illustration of the logical symbol and corresponding Markov model.
(b) The probability of $V_\mathrm{out}$ with blue and green shading corresponding to different cross-voltages. The left (right) panel illustrate 
the dynamics of charging (discharging) the inverter, with input voltages switched at $t=0$. 
(c) Trade-off between first passage time certainty and heat dissipation for 
 discharging and charging over a range of $V_\mathrm{d}$ with $\alpha=0.05$, with the 
shaded region forbidden by the thermodynamic uncertainty relation. 
(d) The same trade-off as in (c) for the charging process shown as a 
function of the probability of a correct gate output in the steady-state
with varying accuracy thresholds $\alpha\in[0, 0.05, 0.1, 0.2]$. 
}
\label{fig:not_gate}
\end{figure}

We start by considering the operation of a single inverter or NOT gate (Fig. \ref{fig:not_gate}a), 
which takes an input binary signal $X$ and outputs its logical inverse $Y$
according to the mapping
\ba
X=
\begin{cases}
0, & V_\mathrm{in}=0 \\
1, & V_\mathrm{in}=V_\mathrm{d} \\
\end{cases}, \ \ 
Y=
\begin{cases}
0, & V_\mathrm{out}\le\alpha V_\mathrm{d} \\
1, & V_\mathrm{out}\ge(1-\alpha) V_\mathrm{d} \\
\emptyset, & \mathrm{otherwise} \\
\end{cases}
\ .
\ea
In the deterministic limit, when the input is $X=1$, current through the P-type transistor is inhibited,
bringing the capacitor gate into effective contact with only the source
reservoir with $Y=0$. 
Conversely, when the input voltage is $X=0$, current is inhibited in 
the other transistor and the capacitor is connected to the drain reservoir, 
with $Y=1$. 
All calculations performed for the single inverter are obtained using numerically exact time evolution, through a Pad\'e approximation \cite{al2010new} with a truncated Hilbert space of $16C_\mathrm{g}(V_\mathrm{d}+4)$, and verified using kinetic Monte Carlo
simulations following the Gillespie algorithm \cite{gillespie1976general}. For calculations where simulation times are not explicitly shown, 
we use time steps distributed logarithmically up to  $10^{16}\beta\hbar$.

Figure~\ref{fig:not_gate}b shows the time dependent response of the 
inverter to an alternating input voltage, with the left and right panels corresponding 
to setting $X=0$ and $X=1$, respectively at $t=0$. 
Lower cross-voltages $V_\mathrm{d}$ require less electron accumulation
in the capacitor gate, and correspondingly fluctuations in the gate output are significantly larger for lower cross-voltages. 
These small accumulations and large fluctuations lead to response times that 
are orders of magnitude faster at low cross-voltages, highlighting a trade-off
between output certainty and characteristic response time. 
While the steady-state statistics of the charged and discharged inverter are symmetric, 
we observe that the dynamics are not. As accentuated at larger $V_\mathrm{d}$, 
the capacitor discharging happens rapidly, while charging occurs relatively slowly. This difference can be understood energetically.
When discharging the capacitor, its occupation regulates the voltage in such 
a way that discharging becomes energetically more favorable as the gate empties. 
In the opposite direction, the accumulation of electrons becomes more energetically unfavorable as 
the gate charges, causing an exponential slowing of current into the gate as a function of time. Additionally, this leads to circulation of electrons between the transistors and 
capacitor when charging, while electrons move directly from the capacitor out of the inverter
when discharging. The functionally unnecessary transitions caused by this circulation
cause slower operation times during the loading process. 

To understand more precisely the interplay between the thermodynamic and operational costs for the inverter, we 
can employ a thermodynamic uncertainty relation \cite{pal_thermodynamic_2021}
\ba\label{eq:fd}
\chi_\mathrm{p}^2=\frac{\langle\tau_\mathrm{p}\rangle^2}{\langle\delta \tau_\mathrm{p}^2\rangle}\le\frac{\langle Q\rangle}{2qV_\mathrm{T}} + 1
\ea
where brackets indicate an trajectory ensemble average, 
$\tau_\mathrm{p}$ is the first passage time to an output voltage
passing the accuracy threshold $\alpha$,
and $\delta x=(x-\langle x\rangle)$, and $Q$ is the the heat dissipated over a trajectory. Explicitly, $\tau_\mathrm{p}$ is calculated from an initial state $X=0$ (or $X=1$), and the final absorbing boundary condition corresponds to a specific cumulative electronic current into (or out of) the capacitor gate  \cite{sharpe2021nearly}. For a specific trajectory, the heat is given as 
\ba
Q = \sum_{k=1}^{N} \ln \frac{W_{x_kx_{k-1}}}{W_{x_{k-1}x_k}}
\ea
where the sum is over steps in the trajectory, and  $x_k$ the state of the system at step $k$ \cite{esposito_three_2010}. 

The thermodynamic uncertainty relation is a general result from stochastic thermodynamics, valid for any Markovian jump process. It states that the certainty in the first passage time, $\chi_\mathrm{p}$, is bounded from above by the heat dissipated over a trajectory, $Q$  \cite{gingrich2017fundamental}.
The bound relates the minimum thermodynamic cost for a given desired certainty in the first passage time. While the bound was originally formulated for a system in steady state, here we apply a finite-time version \cite{pal_thermodynamic_2021}. 

In practical terms, higher certainty in operation time allows for processing input bits at higher rates. For some acceptable probability of error, lower first passage time certainty $\chi_\mathrm{p}$ requires waiting longer to be sure each operation has successfully completed. Conversely, higher first passage time certainty does not require such waiting time, which results in lower heat dissipation due to the shorter overall operation time.

Figure~\ref{fig:not_gate}c shows how this bound depends on the
cross-voltage $V_\mathrm{d}$ for accuracy threshold $\alpha=0.05$. We observe that the bound is not saturated across all $V_\mathrm{d}$ and for both charging and discharging, which implies that the operation of the NOT gate may not be limited by thermodynamic constraints, but rather by its design. Theoretically, it should be possible to design a NOT gate that results in lower heat dissipation for the same level of first passage time certainty, possibly using a different architecture.

In Fig.~\ref{fig:not_gate}d, we vary the cross-voltage and plot the trade-off between first passage time certainty and heat dissipation, as a function of the probability of an error in the output at long times $P_\mathrm{error}=1-\langle Y \rangle_{X=0}$, where the 
ensemble average is over trajectories with the specified input.
We additionally show multiple curves corresponding to differing values of the signal accuracy threshold $\alpha$ and observe that loosening $\alpha$ moves the curves significantly towards the bound by dissipating less heat.
This trend suggests thermodynamically optimal 
operation at higher probabilities of correct outputs as $\alpha$
increases, with similar but less pronounced effects not shown for the discharging process.

\section{Memory Device} \label{memory}

\begin{figure}[b]
\includegraphics[width=8.6 cm]{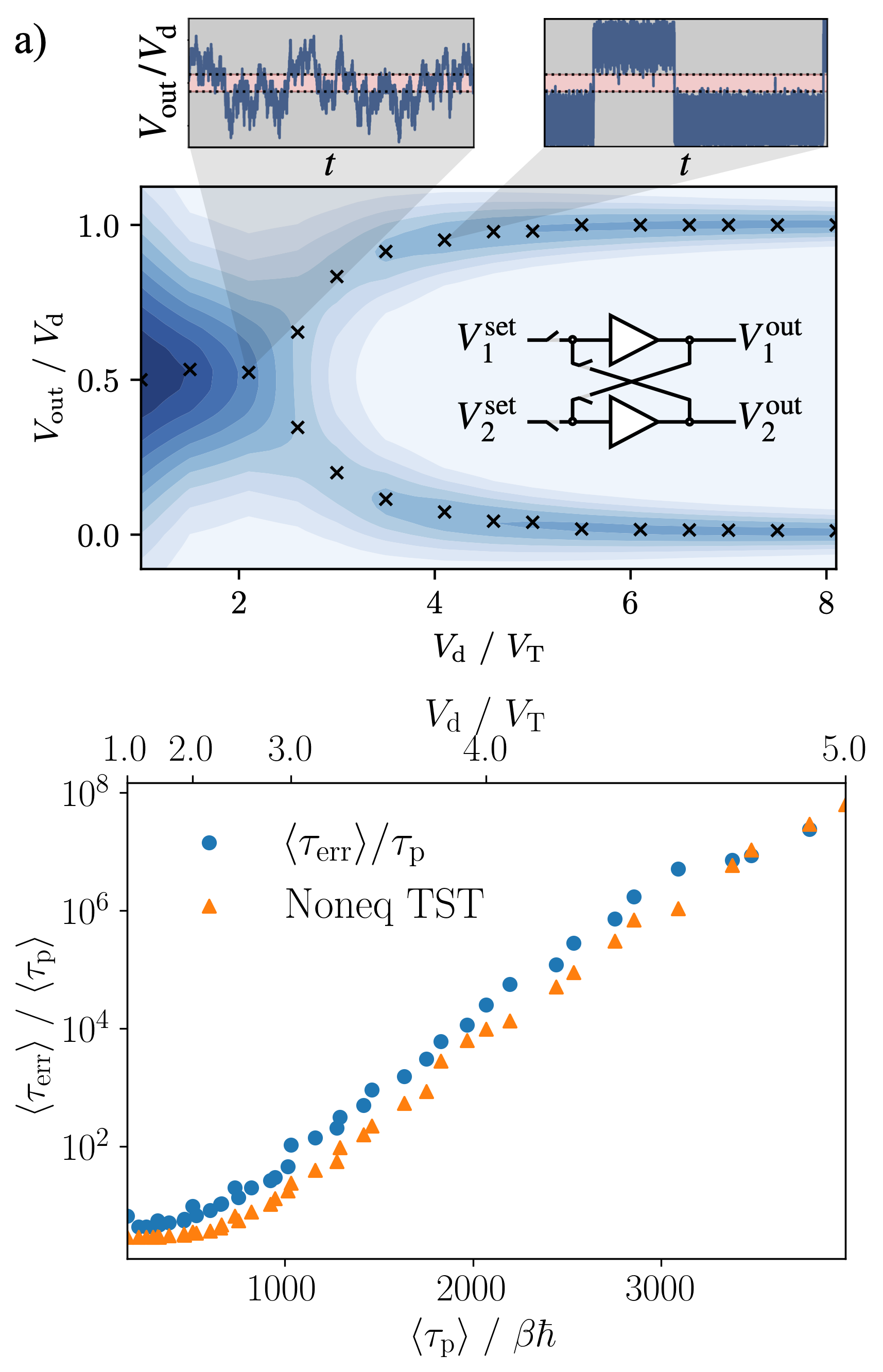}
\caption{
Characterization of the memory device's behavior and 
adherence to thermodynamic speed limits.
(a) Probability of $V_\mathrm{out}$ as a function of $V_\mathrm{d}$, with crosses locating points of maximum probability at each $V_\mathrm{d}$ illustrating the onset of bistability. 
Example trajectories are shown for before and after the onset of bistability. 
(b) Mean time for a memory error as a function of the inverter relaxation time, 
controlled by increasing $V_\mathrm{d}$, as indicated on the top x-axis label.
}
\label{fig:mem_device}
\end{figure}

Next we consider a static random access memory (SRAM) device built by coupling two inverters, as shown in the inset of Fig.~\ref{fig:mem_device}a. 
This device operates using so-called flip-flop circuitry, meaning it exhibits a bistable steady-state, which is a dynamical consequence of a pitchfork bifurcation.
The inverter's state can be set by switching on $V_\mathrm{set}^1$, employing feedback from $V_\mathrm{out}^1$ to $V_\mathrm{in}^2$. 
Here, we will focus on memory maintenance, which can be reliably achieved at sufficiently large cross-voltages by switching both setting voltages off and both feedback loops on. 
To simulate the memory device, we perform an approximate evolution using a fourth-order Runge-Kutta scheme acting on a truncated Hilbert space using a time step of $\Delta t=\beta\hbar/10$ until a final time of $t_\mathrm{f}\approx10^{6}\beta\hbar$. We additionally evaluate the dynamics using kinetic Monte Carlo simulations.

Figure~\ref{fig:mem_device}a shows the steady-state probability of observing an output 
voltage $V_\mathrm{out}=V^\mathrm{out}_1$. The requisite bistability for memory storage
arises at $V_\mathrm{d}/V_\mathrm{T}\approx 2.5$ where the cross-voltage 
overcomes the effects of thermal fluctuations. Notably the bistability is a unique consequence of the nonequilibrium driving, which disappears in the absence of a finite cross-voltage.
At finite $V_\mathrm{d}$, the degeneracy of the steady-state manifests as a spontaneous switching of $V_\mathrm{out}$ as a function of time, illustrated in Fig.~\ref{fig:mem_device}a.
For large $V_\mathrm{d}$ we observe $V_\mathrm{out}/V_\mathrm{d}$ is localized near 0 or 1 for time scales much larger than 
individual inverter operation time scales $\langle\tau_\mathrm{p}\rangle$, indicating persistent memory storage. 
At long times, however, output voltage is stochastically inverted, corrupting the memory storage.

We define $\tau_\mathrm{err}$ as the time required for a memory device, initialized in one of the 
bistable states, to experience a bit flip memory error. 
Figure~\ref{fig:mem_device}b shows the average time required for a bit flip to occur $\langle\tau_\mathrm{err}\rangle$
~\footnote{$\langle\tau_\mathrm{err}\rangle$ is computed by initializing the gate in one of the bistable states 
then using absorbing boundary conditions as is standard \cite{sharpe2021nearly} to compute the 
probability distribution of first passage times into the inverse state.
For practical reasons, because the distribution of first passage times takes the form
$P(\tau_\mathrm{err})\sim e^{-k\tau_\mathrm{err}}$, for large $V_\mathrm{d}$,
we use calculations at intermediate times to compute $k$ from which we calculate 
$P(\tau_\mathrm{err})$ and $\langle\tau_\mathrm{err}\rangle$.
}
as a function of the characteristic time of a single inverter.
We observe that the rate of memory error occurrences decreases exponentially with respect to $\langle\tau_\mathrm{p}\rangle$, 
thus increasing the average memory stability time by roughly five orders of magnitude, from about $100$ ns to $20$ ms for $V_\mathrm{d}/V_\mathrm{T} =2$ to 5.

The efficiency of this memory device can be quantified using a nonequilibrium version of transition state theory \cite{hanggi_tst,chandler_statistical_1978,kuznets2021dissipation}. Transition state theory bounds the rate of a transition between two metastable states using the stationary distribution $P_\mathrm{ss}(V_\mathrm{out})$ and an uncorrelated estimate of the time to cross a dividing surface between the two states. Equivalent to Kramers' theory, the rate is estimated by the probability of a rare fluctuation, in this case a fluctuation of the output voltage of one of the NOT gates of the memory storage device being equal to $V_\mathrm{d}/2$. Taking the dividing surface to be $V_\mathrm{out}=V_\mathrm{d}/2$, and the typical time to relax from the top of the barrier as $\langle \tau_\mathrm{p} \rangle$, a nonequilibrium transition state theory estimate for $\langle\tau_\mathrm{err}\rangle$ is
\ba\label{eq:tst_model}
\langle\tau_\mathrm{err}\rangle\gtrsim \langle\tau_\mathrm{p}\rangle\frac{P_\mathrm{ss}(V_\mathrm{out}/V_\mathrm{d}\le\alpha)}{P_\mathrm{ss}(V_\mathrm{out}/V_\mathrm{d}=1/2)}
\ea
which is shown in Fig.~\ref{fig:mem_device}b. Here the steady-state has been evaluated numerically with $\alpha=0.4$, but because the time-scale for crossing the barrier is significantly larger than relaxation times within each well, the results are similar for $\alpha \in [0, 1/2]$. The nonequilibrium transition state theory provides a very accurate estimate of the memory time, reflecting the likelihood of observing a fluctuation of $V_\mathrm{out}=1/2$ as becoming exponentially unlikely with increasing $V_\mathrm{d}$ in accord with recent large deviation function analysis \cite{freitas2022reliability}. 
Since the transition state theory estimate corroborates the rate of transition between the two bistable states for the potential energy landscape at a given $V_\mathrm{d}$, we conclude that the energy pumped into the SRAM device is efficiently directed into preserving the memory state, rather than spent on extraneous fluctuations which would result in more frequent bit flip errors.

\section{Clock} \label{clock}

\begin{figure}
\includegraphics[width=8.6cm]{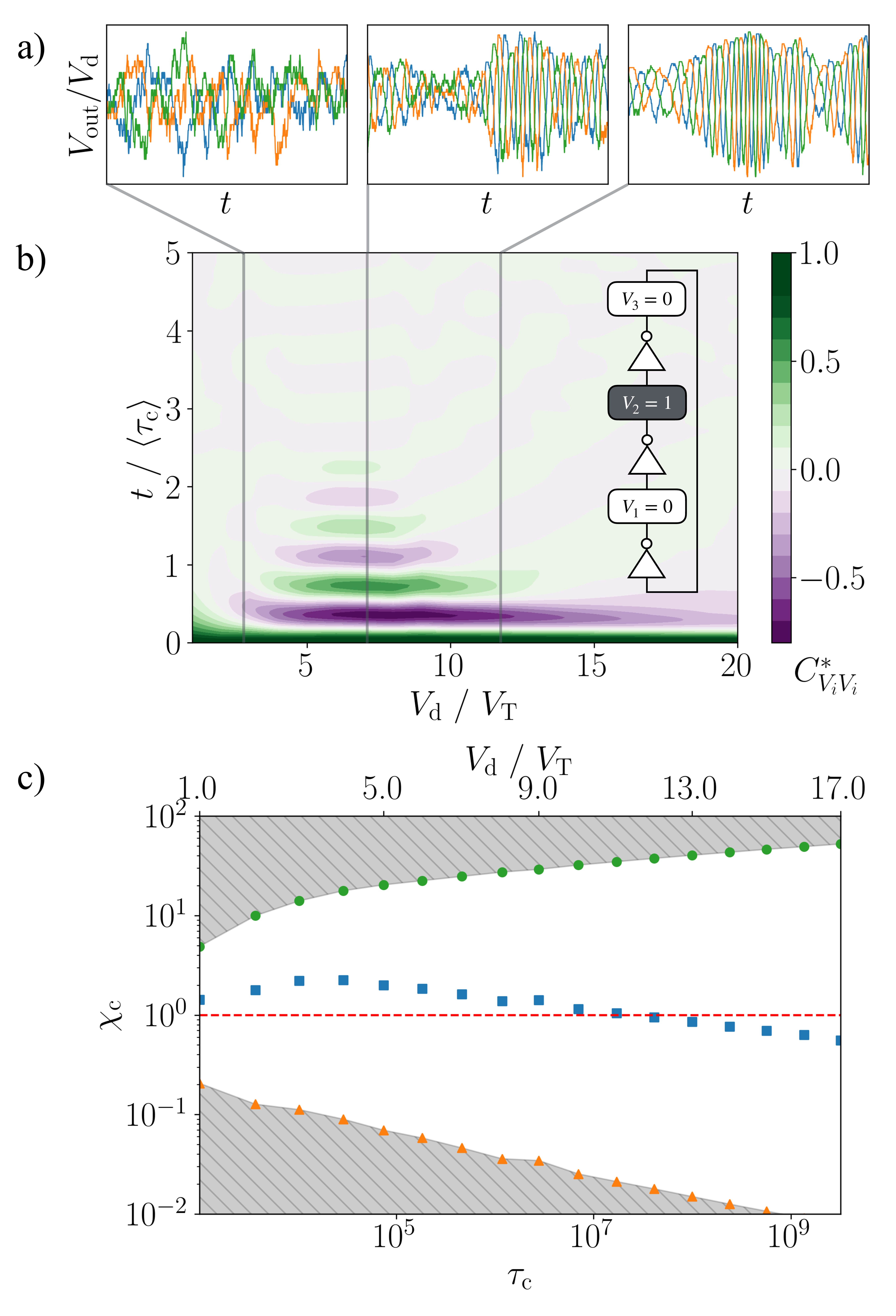}
\caption{
Adherence to thermodynamic bounds 
of the logical clock. 
(a) Example trajectories at $V_\mathrm{d}=[2V_\mathrm{T}, 7 V_\mathrm{T}, 12 V_\mathrm{T}]$, 
with each curve representing how the output voltage of one of the inverters
evolves. 
(b) Rescaled voltage autocorrelation function 
$C_{V_{i},V_{i}}^*(t)=C_{V_{i},V_{i}}(t)/C_{V_{i},V_{i}}(0)$ as a function of applied cross-voltage. 
(c) The certainty in clock operation time $\chi_\mathrm{c}$ (blue squares) as a function 
of average clock cycle time $\langle \tau_\mathrm{c}\rangle$ (bottom axis) 
and applied cross-voltage (top axis),
with the shaded regions indicating the forbidden regions from the uncertainty relations. 
The red dashed line indicates where
$\langle \tau_\mathrm{c}\rangle^2 = \langle \delta\tau_\mathrm{c}^2\rangle$.
}
\label{fig:clock}
\end{figure}

An uneven number of inverters coupled sequentially in a ring creates a system with a frustrated steady-state, because all inverters cannot simultaneously output the logical inverses of their inputs. 
This frustration causes cyclic oscillations, whose period is controlled by inverter operation times, making the device operate as a clock under deterministic conditions and providing an example of circuitry with non-trivial functionality. 
To simulate the dynamics of such a clock, we perform kinetic Monte Carlo
simulations. 
We define the time for the clock to undergo a single cycle $\tau_\mathrm{c}$ 
as the time for the output, $Y$, to cycle from $1-\alpha$ 
to $\alpha$ and back again to $1-\alpha$, using $\alpha=0.4$. Example trajectories are shown in Fig. \ref{fig:clock}a.
All results are averaged over simulations containing at least 50,000 clock cycles. 

In Fig.~\ref{fig:clock}b, we show the output voltage 
autocorrelation function $C_{V_i,V_i}(t)=\langle \delta V_i(0)\delta V_i(t)\rangle$, as a function of time and cross-voltage. 
At low cross-voltages, the three output voltages evolve nearly 
independently, with $C_{V_i,V_i}(t)$ revealing exponential correlations. 
Above $V_\mathrm{d}\approx3 V_\mathrm{T}$ persistent oscillations emerge
but are damped by the stochasticity of the evolution. 
We find a maximal persistence in the autocorrelation function 
oscillation at $V_\mathrm{d}\approx7 V_\mathrm{T}$, 
where the clock undergoes approximately 3.5 cycles. In this region
oscillations are persistent for long times and the fluctuations in 
$\tau_\mathrm{c}$ are small relative to the mean cycle time. 
Above $V_\mathrm{d}\approx10 V_\mathrm{T}$, the autocorrelation exhibits
oscillation for only 1/2 of a cycle because, while oscillations are 
persistent for long times, the fluctuations in $\tau_\mathrm{c}$ are 
large relative to the mean cycle time as anticipated from Fig.~\ref{fig:not_gate}c and the single NOT gate. In this regime, we find the 
cycle time to be inversely proportional to to voltage output amplitude,
which evolves stochastically. 

We define the certainty in the cycle time as
$\chi_\mathrm{c}=\sqrt{\langle \tau_\mathrm{c}\rangle^2 / 
\langle \delta\tau_\mathrm{c}^2\rangle}$, and plot this 
for $V_\mathrm{d}\in[1 V_\mathrm{T},20 V_\mathrm{T}]$ in Fig.~\ref{fig:clock}b. 
The red dashed line indicates where the cycle time's fluctuations 
are equal to its mean, where we find a narrow range of cross-voltages where there is reliable cycling. 
The first passage time thermodynamic uncertainty relation expressed in Eq.~\ref{eq:fd} can be applied to give an upper bound on this quantity 
\ba\label{eq:clock-tur-upper}
\chi_\mathrm{c}^2=\frac{\langle\tau_\mathrm{c}\rangle^2}{\langle\delta \tau_\mathrm{c}^2\rangle}\le\frac{\langle Q\rangle}{2qV_\mathrm{T}}
\ea
where $\langle Q \rangle$ is the average heat dissipated over a cycle. As in the single NOT gate, $\tau_\mathrm{c}$ measures the time to reach a specific cumulative electronic current into or out of the capacitor gate.

Similarly, the dissipation-time uncertainty relation \cite{falasco2020dissipation} relates the rate of heat dissipation with the mean time to complete a process. It can be applied for the clock cycle time to yield a lower bound, 
\ba\label{eq:clock-tur-lower}
\langle\tau_\mathrm{c}\rangle &\ge \left(\beta\langle \dot{Q} \rangle\right)^{-1}\\
\chi_\mathrm{c}^2 &\geq \frac{\langle\tau_\mathrm{c}\rangle}
{\beta \langle\dot{Q}\rangle \langle\delta\tau_\mathrm{c}^2\rangle} \, ,
\ea
where $\langle\dot{Q}\rangle$ is the average rate of heat dissipation in the steady-state.

Fig.~\ref{fig:clock}b exhibits the upper (green circles) and lower bounds (orange triangles) as shaded regions. The upper bound is best saturated when $\chi_\mathrm{c}$ is maximized around $V_\mathrm{d} \approx 3 V_\mathrm{T}$. This occurs for a large enough cross-voltage that the coupled inverters exhibit bistability, but not so large that the fluctuations in the time to charge or discharge each gate are larger than the overall cycle time, preventing reliable cycle behavior.

\section{Clock Control} \label{control}
\begin{figure}
\includegraphics[width=8.6cm]{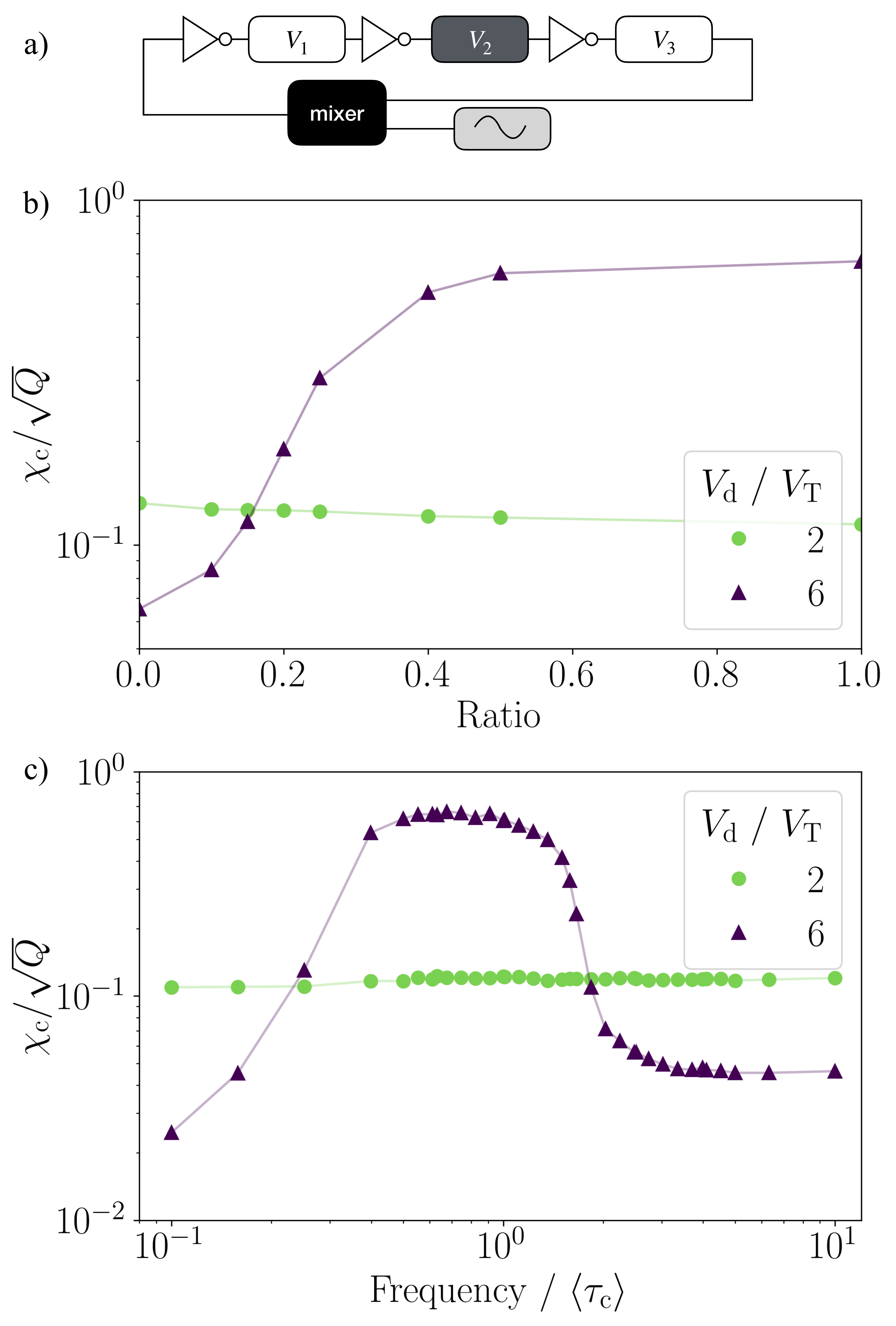}
\caption{
Improvement in reliability of the logical clock cycle time with external control. 
(a) Control mechanism, where the input signal is mixed between the output of the rightmost NOT gate $V_3$ and the external sinusoidal controller.
(b) Certainty in clock operation time $\chi_\mathrm{c}$ as a function of mixing ratio of signal from $V_3$ or the external controller. 
(c) Certainty in clock operation time $\chi_\mathrm{c}$ as a function of the frequency of the external controller. 
}
\label{fig:clock-control}
\end{figure}

To expand the range of reliable cycle times, we propose a simple control mechanism acting on the three-gate circuit. In Fig. \ref{fig:clock-control}a we show a schematic diagram: we insert a sinusoidal oscillator yielding a voltage 
\ba
V_\mathrm{ext}=\frac{V_\mathrm{d}}{2}\sin\left(\frac{2\pi f}{\tau_\mathrm{c}} t\right) + \frac{V_\mathrm{d}}{2}
\ea
whose frequency $f$ we can control with respect to the clock cycle time $\tau_\mathrm{c}$. The signal into the leftmost inverter in the circuit is then given by a linear combination of the output of the rightmost inverter $V_3$ and the external sinusoidal oscillator 
\ba
V_\mathrm{in}=rV_\mathrm{ext} + (1-r)V_3
\ea
where $r\in[0,1]$ denotes the mixing ratio. Here, $r=0$ denotes an undriven system with $V_\mathrm{in} = V_3$, and $r=1$ denotes a fully driven system with signal completely from the sinusoidal oscillator, $V_\mathrm{in} = V_\mathrm{ext}$.

Although this controlled system does not satisfy a TUR because it is not time homogeneous, we can still use the cycle time precision $\chi_\mathrm{c}$ defined previously as a metric for the efficiency of the clock.
We find that this simple control mechanism can improve the reliability of the clock cycling in Fig. \ref{fig:clock-control}b. 
At low driving voltage ($V_\mathrm{d}=2 V_\mathrm{T}$, green circles), the fluctuations in the signal are large enough that external control has little effect. 
With sufficiently high driving ($V_\mathrm{d} = 6 V_\mathrm{T}$, purple triangles), however, we see that precision improves with increasing contribution from the oscillator from left  to right. 
All data in this figure were generated with oscillator frequency equal to $1/\langle \tau_\mathrm{c} \rangle$.

In Fig. \ref{fig:clock-control}c, we vary the frequency of the oscillator relative to the average cycle time $\langle \tau_\mathrm{c} \rangle$ of the uncontrolled circuit for a fixed ratio of $r=0.5$. As before, the external controller has no effect at low driving voltage $V_\mathrm{d}$. At higher driving, the clock precision is maximized when the oscillator frequency is slightly lower than $1/\langle \tau_\mathrm{c} \rangle$, which we attribute to the oscillator acting roughly on resonance with the natural frequency of the circuit. The distribution of $\tau_\mathrm{c}$ is asymmetric, so its average $\langle \tau_\mathrm{c} \rangle$ is slightly higher than the peak of the distribution. Above and below the natural frequency $1/\langle \tau_\mathrm{c} \rangle$, the external oscillations act functionally as noise, to which the system responds unfavorably.

\section{Conclusion} \label{conclusion}
We have used a Markovian model for realistic logical inverters in the regime of thermodynamic computing to explore the interplay between operating characteristics, like accuracy and time, and thermodynamic properties, particularly heat dissipation.
Our results demonstrate the theoretical limits of CMOS circuits using bounds derived from stochastic thermodynamics. 
As we have shown, this provides a framework for simultaneously exploring the fundamental behavior of noisy computational circuits, characterizing their optimality, and using the gained insight to propose more efficient operating procedures and circuits. 
We expect this work will provide a foundation for future work towards understanding and designing efficient thermodynamic computers. 
Current techniques should allow for the improvement of thermodynamic efficiencies by adapting principles from optimal control theory to operational control schemes \cite{das2019variational,lucero2019optimal,engel2022optimal,boyd2022shortcuts,guery2022driving,wimsatt2021refining, zhong_limited-control_2022, blaber_optimal_2023} and by exploring the effects of circuit layout \cite{wolpert2020thermodynamics}.
To extend this approach to larger circuits, more scalable simulation techniques, such as tensor network methods \cite{orus2014practical,helms2020dynamical,causer2022optimal,banuls2019using,strand2022using}, can be adapted. 

The source code for the calculations done and all data presented in this work 
are openly available on Zenodo at https://doi.org/10.5281/zenodo.12708810
 \cite{code}.

\section*{Acknowledgments}
We thank Gavin Crooks for useful discussions. This work was supported by NSF Grant CHE-1954580.

\bibliography{main}
\end{document}